\documentstyle[aps,prl,twocolumn,epsf,rotate]{revtex}
\newcommand{\rr}{{\bf r}}

\begin{document}
%% The following two lines should be there when using 'twocolumn'.
\twocolumn[\hsize\textwidth\columnwidth\hsize\csname
@twocolumnfalse\endcsname

\title{Disorder and interactions in quantum Hall
ferromagnets: effects of disorder in Skyrmion physics}

\author{Jairo Sinova$^{a}$,
A.H. MacDonald$^{a}$
S. M. Girvin$^{c}$}

\address{
         $^a$Department of Physics, University of Texas, Austin, TX 78712, 
         USA\\
         $^b$Department of Physics, Yale University, New Haven, CT 06520
         USA}
\maketitle

\begin{abstract}
We present a Hartree-Fock study of the competition
between disorder and interactions in quantum Hall ferromagnets near $\nu=1$.
We find that the ground state at $\nu=1$ evolves with increasing interaction strength
from a quasi-metallic paramagnet, to a partially spin-polarized ferromagnetic Anderson insulator,
and to a fully spin-polarized ferromagnet with a charge gap.
Away from $\nu=1$, the ground state evolves from a
conventional Anderson insulator, to a conventional quasiparticle glass,
and finally to a ferromagnetic Skyrmion quasiparticle glass.
These different regimes can be measured in low-temperature transport and NMR experiments.
We present calculations for the NMR spectra in different disorder regimes.
\end{abstract}

%\begin{keyword}
%Quantum Hall Ferromagnets, Skyrmions, Disorder Systems
%\PACS 73.40.Hm, 76.60-k, 67.80.Jd, 73.20.Fz, 76.60.Cq
%\end{keyword}

%% The following line should be there when using 'twocolumn'.
\vskip2pc]

\section{Introduction}
At Landau level filling factor $\nu=1$ in the clean limit the ground state of
a two-dimensional electron system (2DES) is a ferromagnet with a charge gap \cite{general}.
Away from $\nu=1$ the ground state is populated by the elementary charge
excitations which are topologically charged spin textures (Skyrmions) containing
more than one single spin flip.  The existence and stability of such
excitations has been verified experimentally through
nuclear magnetic resonance (NMR) experiments \cite{Barrett}.
Recent experiments have indicated that disorder (even in the weak limit)
also plays a role in the observed behavior in a more detailed study of the
NMR spectra at low temperatures \cite{Barrettcondmatt,JairoNMR}.
The competition between interactions and disorder
in the study of quantum Hall ferromagnets has often, but not always \cite{Green,Nederveen},
been neglected partially because of the lack of easily manageable techniques which
can deal with both at the same level.

\begin{figure}
%\centerline{\includegraphics[width=7.6cm]{phasediagram.eps}}
\epsfxsize=3.2in
\centerline{\epsffile{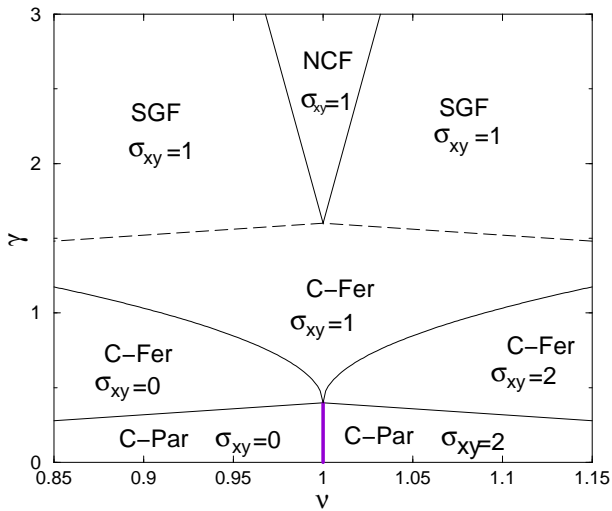}}
\caption{Phase diagram summarizing the results. Here C-Par indicates the
compressible paramagnetic ground state, C-Fer the partially polarized
compressible ferromagnetic ground state, SGF the spin
gapped ferromagnetic ground state, and NCF the non-collinear ferromagnetic ground state.
This phase diagram is qualitative in nature and transition points vs. $\gamma$ should be taken as
upper limits to any realistic phase diagram.}
\label{phasediag}
\end{figure}

Such competition can be studied using the Hartree-Fock (HF)
approximation which has the advantage of being exact in both the non-interacting
and the non-disordered limits \cite{general,jairo}.
We can summarize the results of our model calculations as follows (shown in Fig. \ref{phasediag}).
As the interaction strength is increased relative to disorder at $\nu=1$,
the 2DES ground state suffers a phase transition  from a paramagnetic to
a ferromagnetic state.
For the disorder models we use, the fully polarized state is reached
when the Coulomb energy scale is approximately twice the
Landau-level-broadening disorder energy scale.  Away from $\nu=1$, screening by mobile charges
reduces the importance of disorder and the system reaches
maximal spin-polarization at smaller interaction strengths.
The maximally polarized ground state at moderate interaction strengths is
best described as a glass of localized conventional (Laughlin)  quasiparticles
formed in the $\nu=1$ fully polarized vacuum.  Only for stronger interactions
do we find a phase transition to a state with non-collinear magnetization in which
the localized particles have Skyrmionic character.
We also present NMR spectra calculations at $\nu=1$ in the different disorder regimes
observed.

\section{Formalism}

The HF approximation allows the interplay between disorder and
interactions to be addressed  in an equal footing while
retaining a simple independent-particle picture of the many-body ground state.
In this section, we outline the basic formalism of
HF approximation calculations in the LLL limit. Further details on this
formalism can be found elsewhere \cite{jairo,eric}.

Here we only consider lowest Landau level (LLL) states in expanding the HF Hamiltonian.
Neglecting the frozen kinetic-energy degree of freedom, the Hamiltonian
in second quantization is written as
\begin{equation}
 H= H_I+ H_{dis}+ H_Z\,,
\end{equation}
where $ H_I$ is the normal Coulomb  part of the Hamiltonian,
$ H_{dis}$ is the external disorder part of the Hamiltonian
\begin{eqnarray*}
 H_{dis}&=&\int d\rr\sum_{\sigma}v_{E}(\rr)
\hat\psi^\dagger_\sigma(\rr)\hat\psi_\sigma(\rr)\,,
\end{eqnarray*}
and ${ H}_Z$ is the Zeeman term
\[
H_Z=-\frac{1}{2}g\mu_B \int d\rr \sum_{\sigma \sigma'}
\hat\psi^\dagger_{\sigma'}(\rr') \hat\psi_{\sigma}(\rr')
\vec{\tau}_{\sigma' \sigma}\cdot\vec{B}(\vec r)\,,
\]
with $\sigma=\uparrow,\downarrow$, $v_E$ being 
disorder potential, and $\tau_i$ being the Pauli matrices.
The Zeeman coupling strength is given by $\tilde g=g\mu_B B/(e^2/\epsilon l)$.
For $v_E(\rr)$ we use a white noise distribution so
$\langle \langle v_E(\rr) v_E(\rr')\rangle \rangle
=\sigma^2\delta(\rr-\rr')$. We define the parameter
$\gamma\equiv e^2/(\in \sigma)$ as the ratio of the relative strength of
interactions and disorder broadening.
We chose the commonly used Landau gauge  elliptic theta functions as our LLL basis
to perform Hartree-Fock calculations in this system.

\section{Results}
\begin{figure}
%\centerline{\includegraphics[width=7.6cm]{NMR.eps}}
\epsfxsize=3.2in
\centerline{\epsffile{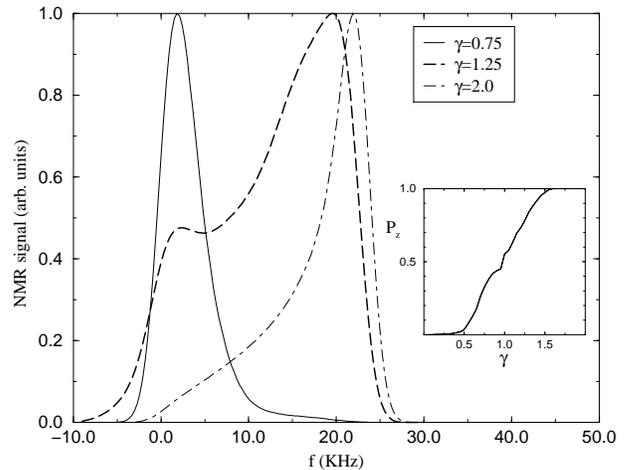}}
\caption{NMR spectrum for $\nu=1$ and $\tilde g =0.015$ at different
$\gamma$'s.
The insert indicates the average magnitude of the local polarization $\rm P_z$ as a function
of $\gamma$.
The sample parameters correspond to ones the used in the experimental
studies of Ref. [3].}
\label{NMRspectra}
\end{figure}

At $\nu=1$ the ground state, in the disorder free limit, becomes a strong
ferromagnet. Such collective behavior has been observed  \cite{Barrett}
in NMR Knight shift experiments in high quality samples.
In figure \ref{NMRspectra} we present the HF theory results
obtained for the dependence of the average magnitude of the spin polarization
(insert) as a function of interaction strength and its corresponding NMR spectra
at the lowest temperatures where there is no motional narrowing of the
spectral line shape \cite{JairoNMR,Yalegrouppapers}. 
The Zeeman coupling strength $\tilde{g}=0.015$
is chosen to match the experimental values. Such NMR intensity spectrum is given by
\begin{eqnarray*}
I(f,\gamma)\propto \int d{\bf r} \rho_N(z)e^{
-\frac{1}{2\sigma^2}(2\pi f -2\pi K_s \rho_e(z) \langle \vec{S}(\rr;\gamma)\rangle)}
\,,
\end{eqnarray*}
with $\sigma=9.34 {\rm ms}^{-1}$ and $K_s\sim 25{\rm KHz}$.
Here $\rho_N(z)$ is the nuclear polarization density
and $\rho_e(z)$ is the electron density envelope function in the quantum well.
The evaluation of such spectra has been outlined elsewhere \cite{JairoNMR,Yalegrouppapers}.
The partially polarized region shows a very distinct spectra which may explain the
existence of anomalous spectra line shapes in certain samples where estimates of the
disorder broadening through mobility measurements \cite{andouemura}
indicate the possibility of being in such regime.

The partially polarized regime can also be studied experimentally by
measuring the transport activation gap.  Provided that weak
Zeeman coupling can be ignored
the Hall conductivity should jump from
$0$ to $2 e^2/h$ at $\nu=1$. In the ferromagnetic state,
the majority-spin extended quasiparticle state will be below
the Fermi level and the Hall conductivity at $\nu=1$
should be quantized at $\sigma_{xy} = e^2/h$.  This spontaneous
splitting is experimentally
accessible and should exhibit interesting non-trivial power law critical
behavior as the ferromagnetic state is entered.
One important feature that our calculations demonstrate is that charge-density fluctuations
at $\nu=1$ do not necessarily require the presence of well-defined Skyrmion
quasiparticles.

\begin{figure}
%\centerline{\includegraphics[width=7.6cm]{Pol_tot_nu_1.25_and_0.75_B_0.1_and_7.0.eps}}
\epsfxsize=3.2in
\centerline{\epsffile{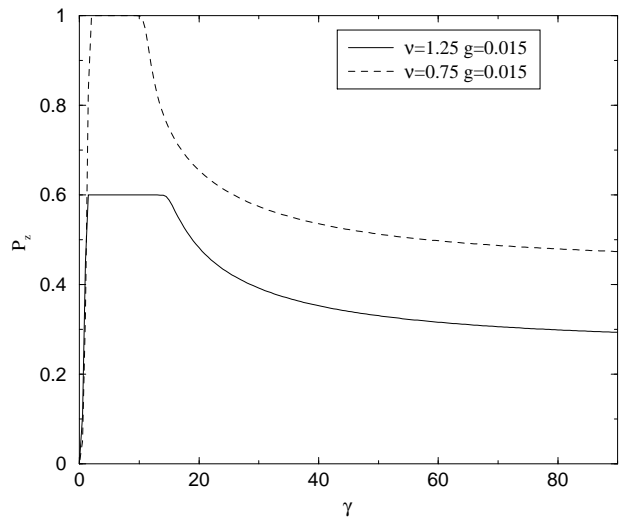}}
\caption{Global polarization phase diagram for a fixed disorder realization.
The transition to a maximally polarized state
(Laughlin quasiparticle glass) occurs at $\gamma\approx \,1.5-2$
for all disorder realizations obtained. The transition from a
Laughlin quasiparticle glass to a Skyrmion glass occurs at $\gamma\approx 15$.
Local polarization and density profiles have been given elsewhere [7].}
\label{phase_diag3}
\end{figure}

In the strong disorder limit
spontaneous spin polarization does not occur at any filling factor near $\nu = 1$.
However, in the large $\gamma$ (clean) limit,  where full polarization is observed at $\nu=1$,
the global polarization decays rapidly with $|1-\nu|$ as observed
experimentally \cite{Barrett}.
The global polarization results for $\nu \ne 1$
in Fig. \ref{phase_diag3}, illustrate how the system interpolates
between these two extrema.
As the interaction strength $\gamma$ is increased from 0 to
2, the behavior mirrors the $\nu=1$ case.  For strong disorder
charge fluctuations  dominate, and small spin polarizations occur
primarily because many single particle orbitals are simultaneously occupied by
both up and down spin electrons. In this regime
charge variation is the main response to disorder
and continues to play an important role at all interaction strengths.
At sufficiently large $\gamma$,
our finite size systems reach a state with the maximum spin
polarization allowed by the Pauli exclusion principle.
This  maximally polarized state is reached earlier than in the case
at $\nu=1$ ($\gamma\sim1.4-1.6$) because, we believe, a larger number of
charged quasiparticles are available to screen the random potential.
At this point the system forms what we refer to as a conventional
quasiparticle glass (CQG).  The
conventional Laughlin quasiparticles are initially localized in the deepest
minima (or maxima for $\nu<1$)  of the
effective disorder potential and as the interaction strength increases,
the charged quasiparticles rearrange themselves locally
into a quasi-triangular Wigner crystal pinned by the strongest of the disorder potential
extrema.
At larger $\gamma$ there is a marked reduction of the
global polarization from its maximally polarized value.
This indicates a transition from a CQG to a Skyrmion glass (here glass is meant
as a non-Bravis lattice arrangement of the Skyrminos).
The point of cross over from the CQG to
the Skyrmion glass, as illustrated in Fig. \ref{phase_diag3},
depends on filling factor and $\tilde g$.
We approximate the dependence of the transition point on $\tilde{g}$ in this regime
by considering a simple model for a single Skyrmion trapped
at a disorder potential extrema.  We approximate its energy by
\begin{equation}
E(K)=U(K-K_0)^2+g^*\mu_B B K+\sigma AK\,,
\label{model}
\end{equation}
where  $K$ is the number of spin flips per Skyrmion, 
and $A$ is a phenomenological parameter.
The first two terms
pin the optimal Skyrmion size in the clean limit \cite{UKmodel_Allan}.
The third term  favors the stronger localization of reduced (small $K$)
Skyrminos close to the potential extrema.
This simple model
gives an estimate of the interaction strength at which $K > 0$ Skyrmions
first become stable
\begin{equation}
\gamma^*=\frac{A}{2K_0U/(e^2/\epsilon \ell)-\tilde{g}}\,.
\label{gammastar}
\end{equation}
At factor $\nu=1.25$,
$U/(e^2/\epsilon \ell)\sim0.014$ and $K_0\sim 1$,
$A\sim 0.1$ \cite{jairo,UKmodel_Allan}. 
From this, one obtains an estimate of $\gamma^*\sim 7$ for the cross over
point from conventional quasiparticles to Skyrmions at $\tilde{g}=0.015$.
This is in reasonable agreement with the actual cross over point
$\gamma^*\sim 10$ (see Fig. \ref{phase_diag3}.
In the clean limit, the Skyrmion system crystallizes in a square
lattice for the filling factors considered here.  (The Skyrmion crystal
is triangular \cite{UKmodel_Allan} for $\nu$ very close to 1.)
These estimates of the
maximum disorder strength at which Skyrmion physics is realized could
be checked by performing NMR experiments in samples where
electron density can  be adjusted through gate voltages.

\section*{Acknowledgements}

This work was supported by the Welch 
Foundation, and NSF grants DMR-9714055 and DMR-9820816.

\end{document}